\documentclass[conference]{IEEEtran}

\usepackage{cite}

\usepackage[pdftex]{graphicx}
\usepackage{epstopdf}
\DeclareGraphicsExtensions{.eps}
\usepackage{amsmath}
\usepackage{amssymb}
\usepackage{algorithmic}
\usepackage{algorithm}
\usepackage{color,colortbl}

\begin{document}

\title{Performance Comparison of Constant Envelope and Zero-Forcing Precoders in Multiuser Massive MIMO}

\author{\IEEEauthorblockN{Alberto Brihuega, Lauri Anttila, and Mikko Valkama}
\IEEEauthorblockA{Tampere University of Technology, Department of Electronics and Communications Engineering,
Tampere, Finland}
\IEEEauthorblockA{Email: alberto.brihuegagarcia@tut.fi, lauri.anttila@tut.fi, mikko.e.valkama@tut.fi}}

\maketitle

\begin{abstract}

In this article, the adoption and performance of a constant envelope (CE) type spatial precoder is addressed in large-scale multiuser MIMO based cellular network. We first formulate an efficient computing solution to obtain the antenna samples of such CE precoder. We then evaluate the achievable CE precoder based multiuser downlink (DL) system performance and compare it with the corresponding performance of more ordinary zero-forcing (ZF) spatial precoder. We specifically also analyze how realistic highly nonlinear power amplifiers (PAs) affect the achievable DL performance, as the individual PA units in large-array or massive MIMO systems are expected to be small, cheap and operating close to saturation for increased energy-efficiency purposes. It is shown that the largely reduced peak-to-average power ratio (PAPR) of the PA input signals in the CE precoder based system allows for pushing the PA units harsher towards saturation, while allowing to reach higher signal-to-interference-plus-noise ratio (SINRs) at the intended receivers compared to the classical ZF precoder based system. The obtained results indicate that the CE precoder can outperform the ZF precoder by up to 5-6 dBs, in terms of the achievable SINRs, when the PA units are pushed towards their saturating region. Such large gains are a substantial benefit when seeking to improve the spectral and energy-efficiencies of the mobile cellular networks. 
\end{abstract}


\IEEEpeerreviewmaketitle

\section{Introduction}
\IEEEPARstart{P}{ower} consumption of the cellular network is commonly recognized as a major concern \cite{Green}. The power-efficiency of radio transmitters, and particularly the involved power amplifiers (PAs), is one key aspect in the total network power consumption  \cite{pa}. Due to the envelope characteristics of the currently used radio access waveforms, power amplifiers need to operate in a relatively linear regime in order not to distort the transmit signals, resulting commonly into low power-efficiency \cite{problemas_papr}. Furthermore, the demands of data hungry users and new services lead to adopting large antenna arrays at the base stations (BS). Such large antenna arrays enable the improvement of the system spectral efficiency linearly proportional to the number of antennas \cite{oportunidades}. In addition to beamforming and multiplexing gains, it has recently been established that large antenna arrays also allow for transmit waveform shaping in such a way that robustness against PA nonlinearities can be achieved \cite{CE1}. Constant envelope type of precoders allow for reducing the peak-to-average power ratio (PAPR) of the corresponding continuous-time PA input signals, while simultaneously providing beamforming and spatial multiplexing benefits, similar to more ordinary linear precoders. This allows to simultaneously address the two main targets of new mobile communications systems, namely improved spectral and energy efficiencies. In the recent literature \cite{CE1,CE2,CE3,CE4,CE5,CE6}, the optimization and characterization of such CE precoders have been addressed, however, none of the existing works provide a comprehensive performance evaluation under realistic measurement-based nonlinear PA units nor comparison with traditional spatial precoders such as zero forcing (ZF). In this paper, the ZF based spatial precoder serves  as the reference precoding technique, due to its simplicity, well known performance and wide-scale utilization in massive MIMO research and development work. Furthermore, the proposed CE based spatial precoder involves setting a constraint on maximum allowed multiuser interference (MUI), as it will be detailed further below. The fact that ZF type of precoder is capable of fully suppressing the MUI makes ZF a better reference than, for example, maximum ratio transmission (MRT) precoder that does not explicitly control MUI. Thus, this ensures that the performance limitations of the reference precoder is due to the nonlinear distortion introduced by the PAs, and not due to any other sources. Thereby, ZF precoder serves well as the reference solution in evaluating and comparing the performance of the CE-based spatial precoder.

In this article, based on above reasoning, we analyze and evaluate the performance of a large-array multiuser (MU) MIMO downlink system, with CE and ZF based spatial precoders. In order to be able to perceive the effects of the PAPR reduction on the realistic system performance, we adopt measurement based nonlinear PA models in every antenna branch of the base-station. For a given transmit power, it is shown that the resulting continuous-time waveforms obtained through the CE precoding exhibit substantially milder inband distortion due to the higher tolerance to the PA nonlinearities, an effect that is more rigorously analyzed and characterized through the experienced signal-to-noise-and-interference ratio (SINR) and bit error rate (BER) at the intended receivers. In case of ordinary ZF precoder, it is also shown that as the transmit power is increased, the inband distortion produced by the nonlinear behaviour of the PAs becomes more and more significant, and eventually will become the main source of interference or distortion experienced in the receivers, and thus becomes the limiting factor to the achievable system performance. The CE precoder does not have such limit as long as the PA units are not fully saturated. Furthermore, in the paper, we describe and implement a computationally efficient optimization approach to obtain the CE precoded antenna samples. Overall, the obtained results reported in the paper show that compared to ordinary ZF precoder, the CE precoder allows for pushing the PA units of a large-array base-station substantially harsher towards saturation, while allowing to reach higher SINRs at the intended receivers. Depending on the transmit sum-power or effective radiated power constraints, the results show that the gain of the CE precoder over ZF can be even up to 5-6 dBs, in terms of the achievable SINRs. Such large gains are a substantial benefit when seeking to improve the spectral and energy-efficiencies of the existing and emerging mobile cellular networks. 

The rest of this paper is organized as follows: In Section II, the basic system model of multi-user massive MIMO downlink transmission is provided, incorporating the construction of the involved CE precoder. Also the PA nonlinearities and other different distortion and interference aspects are addressed. Then, in Section III, the performance evaluation results and their analysis are provided. Finally, Section IV will conclude the work and summarize the main findings.
\section{System Model and CE Precoder}
We assume a large-scale MU-MIMO downlink system where $K$ and $N_{t}$ denote the number of single-antenna simultaneously scheduled users and the number of transmit antennas at the base station, respectively, where $N_{t}>>K$. It is assumed that there is a total symbol-rate transmit sum-power constraint $P_{t}$. Let $\mathcal{S}$ denote the information alphabet and $\mathbf{s} = [s_{1}, s_{2},\cdots , s_{K}]^{T}$ denote the vector of information symbols, where $s_{k} \in \mathcal{S}$ denotes the information symbol intended for the $k$-th user. Furthermore, in case of linear precoding, let $\mathbf{W}_{TX} \in \mathbb{C}^{N_{t} \times K}$ denote the precoding matrix which will be obtained through the ZF principle, and used as the reference method in this paper.  The linear precoded symbols $\mathbf{x} = [x_{1}, x_{2},\cdots , x_{N_{t}}]^{T}$ are obtained as
\begin{equation}
 \mathbf{x} = \mathbf{W}_{TX}\mathbf{s}. 
 \label{linear_precoded_samples}
\end{equation}
The CE precoder, in turn, is a nonlinear mapping from the information symbols $\mathbf{s}$ to the precoded samples $\mathbf{x}$, which we will address explicitly later in this section.

For mathematical tractability, we assume a single-carrier system, as in \cite{CE1,CE2,CE3,CE4,CE5, CE6}, and thus root raised-cosine (RRC) filters are utilized for filtering the precoded and upsampled symbols, generating thus the continuous-time signals $\mathbf{x}(t) = [x_{1}(t), x_{2}(t),\cdots , x_{N_{t}}(t)]^{T}$. Notice that since the CE precoder operates at symbol level, the continuous-time signals are not exactly CE waveforms since the RRC filtering introduces some inherent PAPR increase - even if the precoded symbols $x_n$ have constant envelope. This PAPR increase and the associated sensitivity to PA nonlinearities will be addressed in Section III.

The continuous-time signals then pass through highly nonlinear PAs. In this work, $9$-th order memoryless polynomial models of the form
\begin{multline}
    z_{n}(t) = b_{1,n}x_{n}(t) + b_{3,n}x_{n}(t)|x_{n}(t)|^{2} +  b_{5,n}x_{n}(t)|x_{n}(t)|^{4} \\ + b_{7,n}x_{n}(t)|x_{n}(t)|^{6} + b_{9,n}x_{n}(t)|x_{n}(t)|^{8}
\end{multline}
obtained from RF measurements of a set of actual PAs, are utilized, where $n$ refers to the antenna/PA index. The polynomial coefficients of the different PA units, obtained from the measurements, are all slightly different reflecting the true characteristics and nature of the measured PAs. In general, since the polynomials behave expansively, at large amplitude levels, they are properly clipped to reflect the true saturation levels of the individual PAs.

For simplicity, we assume narrowband fading, and thus, the channel between the $k$-th receiving antenna and the $n$-th transmit antenna can be modeled as a single complex coefficient. The corresponding zero-mean-unit-variance flat-fading Rayleigh multiuser channel matrix is denoted by $\mathbf{H} \in \mathbb{C}^{K \times N_{t}}$. Furthermore, perfect channel state information (CSI) knowledge is assumed at the transmitter.

The well-known ZF precoder coefficients \cite{ZF_c}, utilized in this paper as the reference technique, are obtained by means of the right pseudoinverse of the multiuser channel matrix as
\begin{equation}
\mathbf{W}_{TX}^{ZF} = \mathbf{H}^{H}\left(\mathbf{H}\mathbf{H}^{H}\right)^{-1}
\label{ZF_pre}
\end{equation}
On the other hand, the CE precoded samples are commonly obtained by means of more elaborate optimization frameworks, as discussed in \cite{CE1,CE2,CE3,CE4,CE5, CE6}. In this work, we formulate next a computationally efficient iterative least-mean square (LMS) type of an approach to obtain such CE precoded samples. First, the  CE precoded samples are constrained to have constant envelope such that $|x_{n}| = \sqrt{P_{t}/{N_{t}}}$, therefore, the precoder outputs $x_{n}$ are of the form
\begin{equation}
x_{n} = \sqrt{\frac{P_{t}}{N_{t}}} e^{j\theta_{n}} \:\:,\:\:\: n = 1,\cdots, N_{t}
\end{equation}
Thus, the precoder generates a symbol rate constant envelope signal in every antenna branch, each of them with a certain phase. The expression in (4) also automatically guarantees that the symbol-rate transmit sum-power constraint of $P_t$ is met Then, by neglecting the PA induced distortion in the algorithm development phase, the CE precoded signal received by the $k$-th user can be consequently expressed as
\begin{equation}
y_{k} = \sqrt{\frac{P_{t}}{N_{t}}}\,\sum_{n=1}^{Nt} h_{k,n}e^{j\theta_{n}} + n_{k} \:\:,\:\:\: k = 1,\cdots, K
\label{received_symbol}
\end{equation}
where $n_k$ refers to additive noise while $h_{k,n}$ denotes the channel between the $k$-th user and the $n$-th transmit antenna. The MUI experienced by the $k$-th receiver can be measured as the difference between the actual noise-free received signal and the intended symbol, and can be expressed, in terms of the instantaneous squared value, as
\begin{equation}
\gamma_{k} =  \left|\left(\sqrt{\frac{P_{t}}{N_{t}}}\sum_{n=1}^{Nt} h_{k,n}e^{j\theta_{n}} -  \alpha s_{k}\right)\right|^{2} 
\end{equation}
where $\alpha$ denotes the obtained beamforming gain towards the intended user. Then, the phases of the precoded samples are selected such that the instantaneous power of the MUI over all intended receivers is minimized. Such phase optimization problem reads thus \cite{CE1}
\begin{gather*}
\Theta = [\theta_{1},\theta_{2},\cdots,\theta_{N_{t}}]=\arg \mathop{\min}_{\theta_{n}\in[-\pi,\pi), n=1,\ldots,N_{t}} f(\Theta, \mathbf{s}) \\
f(\Theta, \mathbf{s})  = \sum_{k=1}^{K} \left|\left(\sqrt{\frac{P_{t}}{N_{t}}}\sum_{n=1}^{Nt} h_{k,n}e^{j\theta_{n}} -  \alpha s_{k}\right)\right|^{2} 
\end{gather*}
\begin{equation}
\begin{split}
s.t. \:\:\:\:\: &\left|\left|\mathbf{x}\right|\right|^{2} = P_{t} \\
&\left|x_{n}\right| = \sqrt{\frac{P_{t}}{N_{t}}} 
\end{split}
\end{equation}
where $\theta_{n}$ denotes the phase of the precoded sample for the $n$-th antenna.
\begin{algorithm}
 \caption{LMS-based optimization framework for CE precoder}
 \begin{algorithmic}[1]
 \STATE $\mathbf{\Theta_{1}} = [0, 0,\cdots, 0]^{T}$
 \STATE $threshold$ = +inf.
  \FOR{$n = 1$ to $N_{t}$}
	\FOR{$m = 1$ to $M$} 
 		\STATE $e_{m} = \sum_{k=1}^{K} \left|\left(\sqrt{\frac{P_{t}}{N_{t}}}\sum_{n=1}^{Nt} h_{k,n}e^{j\theta_{n},m} -\alpha s_{k}\right)\right|^{2} $
 		\STATE $\theta_{n,m+1} = \theta_{n,m} + \theta_{LMS,m}(e_{m})$
		\IF{ $e_{m} < threshold$}
			\STATE $threshold = e_{m} $
			\STATE $\theta_{n,opt} = \theta_{n,m}$
		\ENDIF
	\ENDFOR
	\STATE $\theta_{n} =\theta_{n,opt} $
\ENDFOR
\RETURN  $\mathbf{x}_{opt} = \left[\sqrt{\frac{P_{t}}{N_{t}}}e^{j\theta_{1,opt}}, \cdots, \sqrt{\frac{P_{t}}{N_{t}}}e^{j\theta_{N_{t},opt}}\right]^{T}$ 
 \end{algorithmic} 
 \end{algorithm}
 
To solve the optimization problem, we adopt an iterative approach described in Algorithm 1. This computing friendly algorithm consists of $N_{t}\times M$ iterations, where $M$ is a certain prefixed integer value. Intuitively, in every $n$-th iteration, the phase of the symbol at the $n$-th antenna branch is adapted following $M$ sub-iterations of the gradient descent algorithm based on the error signal $e_{m}$, while the phases of the rest of the antenna branches remain fixed. Then, the phase of the $m$-th sub-iteration which resulted in the lowest MUI (denoted by $\theta_{n,opt}$) is assigned to the $n$-th antenna. Then, the algorithm proceeds to the $(n+1)$-th iteration.

In general, if one wants to provide a certain beamforming gain $\alpha$, a constraint on the maximum allowed MUI is required. Then, $\mathbf{\Theta}$ is selected such that $\alpha$ is maximized, while keeping the MUI below the maximum allowed limit. For further details, please, refer to \cite{CE1}.

In order to later compare the system performance with different precoders in a fair manner, specific care is needed in constraining the transmit power. In general, one approach is to assume that the transmit sum-powers under both precoders are constrained identical. It is to be noted, however, that due to the fact that the two precoders may present somewhat different beamforming gains, the equivalent isotropic radiated powers (EIRP) may also be different. Thus, in the following performance evaluations and comparisons, we consider two scenarios, one in which the transmit power is scaled in such a way that both precoders present the same EIRP, and a second scenario assuming that both precoders have the same transmit power, and thus, their EIRPs are different.

In order to constrain the output sum-power of the ZF precoder, one can first consider a general linear precoded signal of the form $\mathbf{x} = \mathbf{W}_{TX}\mathbf{s}$. In order to constrain the sum-power to $P_t$, we first express the covariance matrix of the precoded samples as
\begin{equation}
cov(\mathbf{x}) = \mathbb{E}\left\{\mathbf{x}\mathbf{x}^{H}\right\}  = \sigma^{2}\mathbf{W}_{TX}\mathbf{W}_{TX}^{H}
\end{equation}
where it has been assumed that the data streams are independent from one another and have a covariance $\mathbb{E}\left\{\mathbf{s}\mathbf{s}^{H}\right\}=\sigma^{2}\textbf{I} $. Assuming further that the individual data stream powers are normalized to one, that is  $\sigma^{2}=1$, the total output sum-power constraint can be expressed as
\begin{equation}
\mathbb{E}\left\{||\mathbf{x}||^{2}\right\} = trace\left\{ cov(\mathbf{x})\right\} =  trace\left\{\mathbf{W}_{TX}\mathbf{W}_{TX}^{H}\right\} = P_{t}
\end{equation}
Thus, the transmit sum-power constraint is met if any given precoding coefficients $\mathbf{W}_{TX}$ are normalized by
\begin{equation}
\beta = \sqrt{\frac{P_{t}}{trace\left\{\mathbf{W}_{TX}\mathbf{W}_{TX}^{H}\right\}}}
\end{equation}
 The normalized precoder output thus reads
 \begin{equation}
 \mathbf{x} = \beta\mathbf{W}_{TX}\mathbf{s}
 \end{equation}
 Notice that the CE precoded samples obtained through Algorithm 1 are, by design, automatically fulfilling the transmit sum-power constraint of $P_t$. Notice also that in case of ZF precoder in (3), the above normalization factor represents directly the beamforming gain $\alpha$, while for CE precoder the beamforming gain has to be calculated numerically. Finally, since the beamforming gains of the CE and ZF precoders are generally different, the EIRPs are also different under the given transmit sum-power constraint. Therefore, if performance comparison under fixed EIRP is pursued, an additional sum-power scaling needs to be adopted.
 
In general, the precoder coefficients obtained by means of the ZF principle need to follow the time variations of the MIMO channel, and therefore, its updating rate is dictated by the coherence time of the propagation channel. On the other hand, CE based precoder optimization to obtain the precoded samples must be executed at every symbol instant. The coherence time of the channel can easily be hundreds or thousands of times longer than the symbol duration, thereby, the CE precoder involves substantially larger computing complexity than the linear precoders. Furthermore, the complexity of the ZF based precoder grows linearly with the number of antennas, while that of the proposed CE increases quadratically with the number of antennas. Developing CE precoders with reduced complexity is thus an important future work item for us.
 
 \section{Performance Results and Analysis}
In this section, detailed performance evaluation results and their analysis are presented. As a concrete example, we focus on a $24 \times 4$ MU-MIMO scenario with four single-antenna users. Four data streams, all of 16-QAM symbols, serve as precoder inputs, while the precoder can be either CE or ZF. The CE precoded samples are optimized such that a 20 dB MUI suppression is guaranteed. For the considered MUI suppression, it can be shown that the CE precoder provides 1.9 dB lower beamforming gain than the ZF precoder. In general, the precoders map the data streams into $24$ antenna branches, and the resulting signals then pass through upsampling and RRC filtering stage of order 33 and with 0.4 roll-off factor. Lastly, 24 different clipped $9$-th order memoryless polynomial models, obtained from extensive RF measurements, are adopted in order to model the behavior of the PA units in true array transmitter\footnote{ Lund University Massive MIMO testbed, http://www.eit.lth.se/mamitheme}. Since normalized polynomial models are used, one needs to properly scale and unscale the input and output signals of the PAs, respectively, such that the signals fit in the polynomial range. 

At the receiver side, we evaluate the BER and SINR in order to quantify the quality of the received signals. Noise level at the receiver side is fixed such  that in the absence of in-band distortion, it constitutes the main source of received signal degradation. In the evaluations, we vary the transmit sum-power $P_{t}$ (under fixed noise level), which has an impact on the resulting BER, back-off and SINR, which are presented in the following subsections. The higher the transmission power, the lower the applied back-off is, and therefore, the PAs introduce higher in-band distortion which degrades the BER in case of ZF precoder. In the figures below, we also plot the mean back-off, relative to the 1dB compression point, as a function of the transmission power. The reason to illustrate the mean back-off is because every PA has a slightly different characteristic. 

\subsection{PAPR Distributions with CE and ZF precoders}
\begin{figure}
	\centering
	\includegraphics[width=1\linewidth]{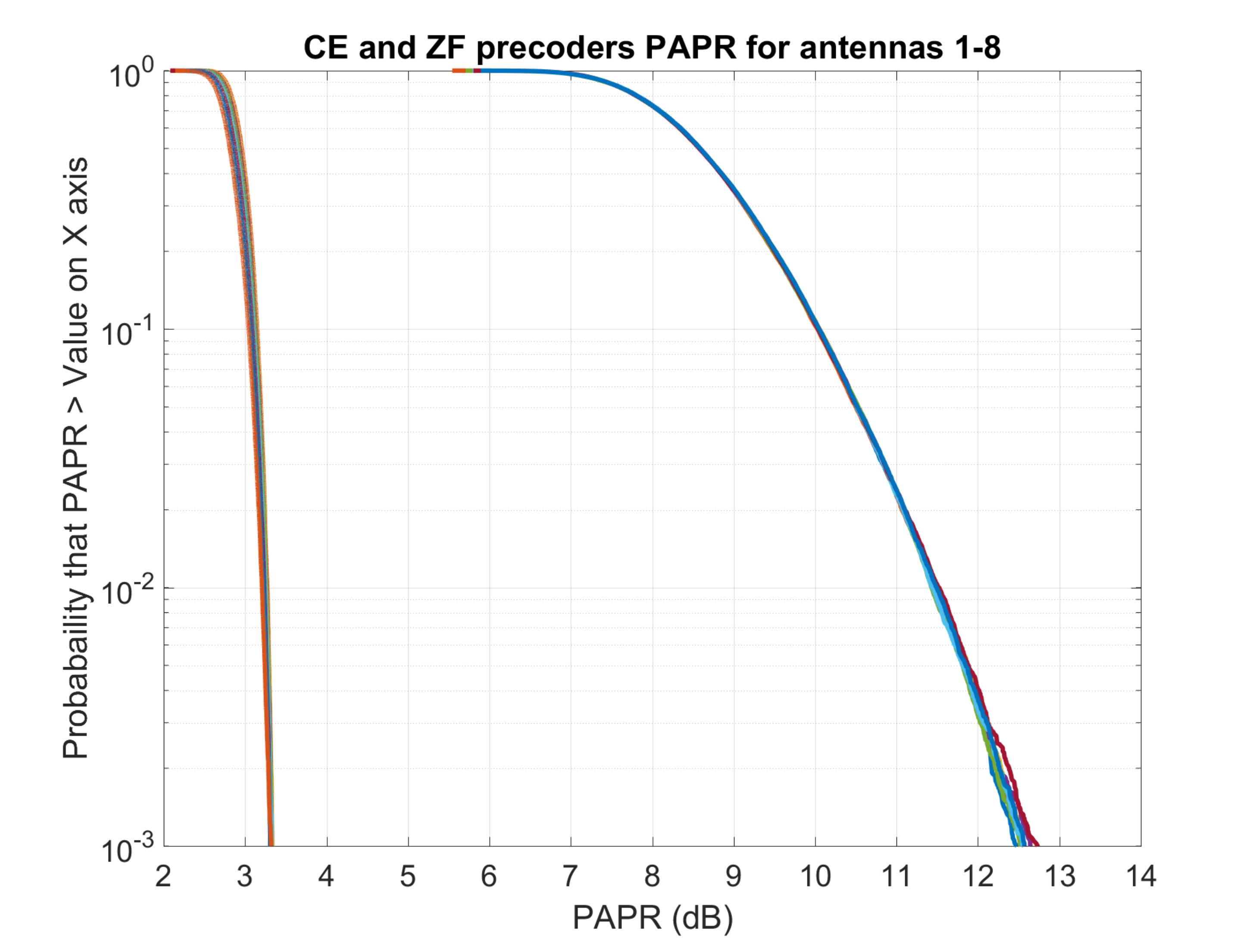}
	\caption{CE and ZF precoder PAPR distributions. The set of curves on the left side corresponds to the CE precoded signals, while the set of curves on the right corresponds to the ZF precoded signals. Only the PAPR CCDFs of the antenna signals 1-8 are shown, while those of 9-24 behave very similarly.}
	\label{fig:PAPR}
\end{figure}

We begin by shortly evaluating and illustrating the complementary cumulative distribution functions of the precoded antenna signals, using both the ZF and CE precoders. The results are shown in Fig. \ref{fig:PAPR}, and illustrate how efficiently the symbol-rate CE precoder is able to reduce the PAPR of the antenna signals despite the RRC filtering stage. While the PAPR of the ZF precoded signals can easily reach a level of 12 dB, the PAPR of the CE coded signals is commonly in the order of 3 dB only. Next we address how this translates to multiuser radio link performance under nonlinear PA units.

\subsection{Multiuser Radio Link BER with Fixed EIRP}
Here we evaluate and analyze the case in which both precoders are scaled such that the EIRP is fixed independent of which precoder scheme is utilized. Note that this corresponds to different transmit sum-powers due to the different beamforming gains that the two precoders are able to provide.

\begin{figure}[t]
	\centering
	\includegraphics[width =0.45\textwidth]{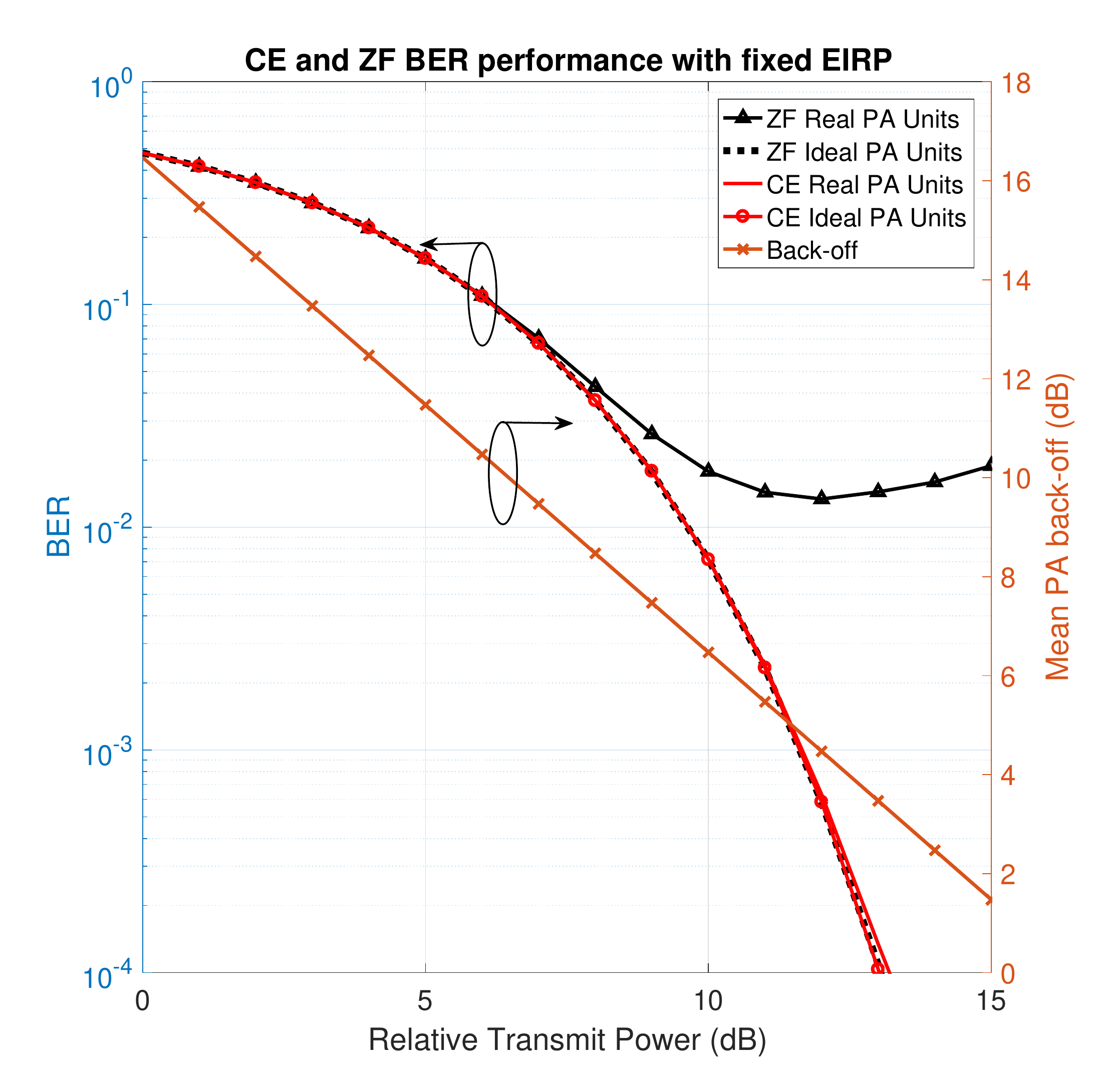}
	\caption{BERs of CE and ZF precoders with (i) \textit{Ideal PA Units} referring to a case with fully linear PAs and with (ii) \textit{Real PA Units} referring to the case with actual measured nonlinear PAs. Precoders are scaled such that both yield the same EIRP. Back-off ranging between 1.5 and 16.5 dB. Relative transmit power of 0 dB corresponds to receiver thermal noise SNR of -1 dB.}
	\label{fig:no_beam_fig}
\end{figure}

The results under fixed EIRP shown in Fig. \ref{fig:no_beam_fig} clearly illustrate how the actual reduction of the PAPR of the CE precoded PA input waveforms allows to push the PAs closer to their nonlinear operation zone for a given in-band distortion. With the ZF precoder, as the transmit power increases, the nonlinear distortion starts to become larger and larger until it constitutes the main source of interference, thus saturating the performance of the ZF precoded system. CE precoder, in turn, exhibits almost ideal performance even when nonlinear PAs are considered. Only at the very highest transmit power levels, the CE precoded system exhibits a very minor BER degradation compared to the fully linear PA case.  
\subsection{Multiuser Radio Link BER with Fixed Transmission Power}
Next, we consider the scenario in which both precoding schemes yield the same transmission power, and thus somewhat different EIRPs. 
\begin{figure}[t]
	\centering
	\includegraphics[width = 0.45\textwidth]{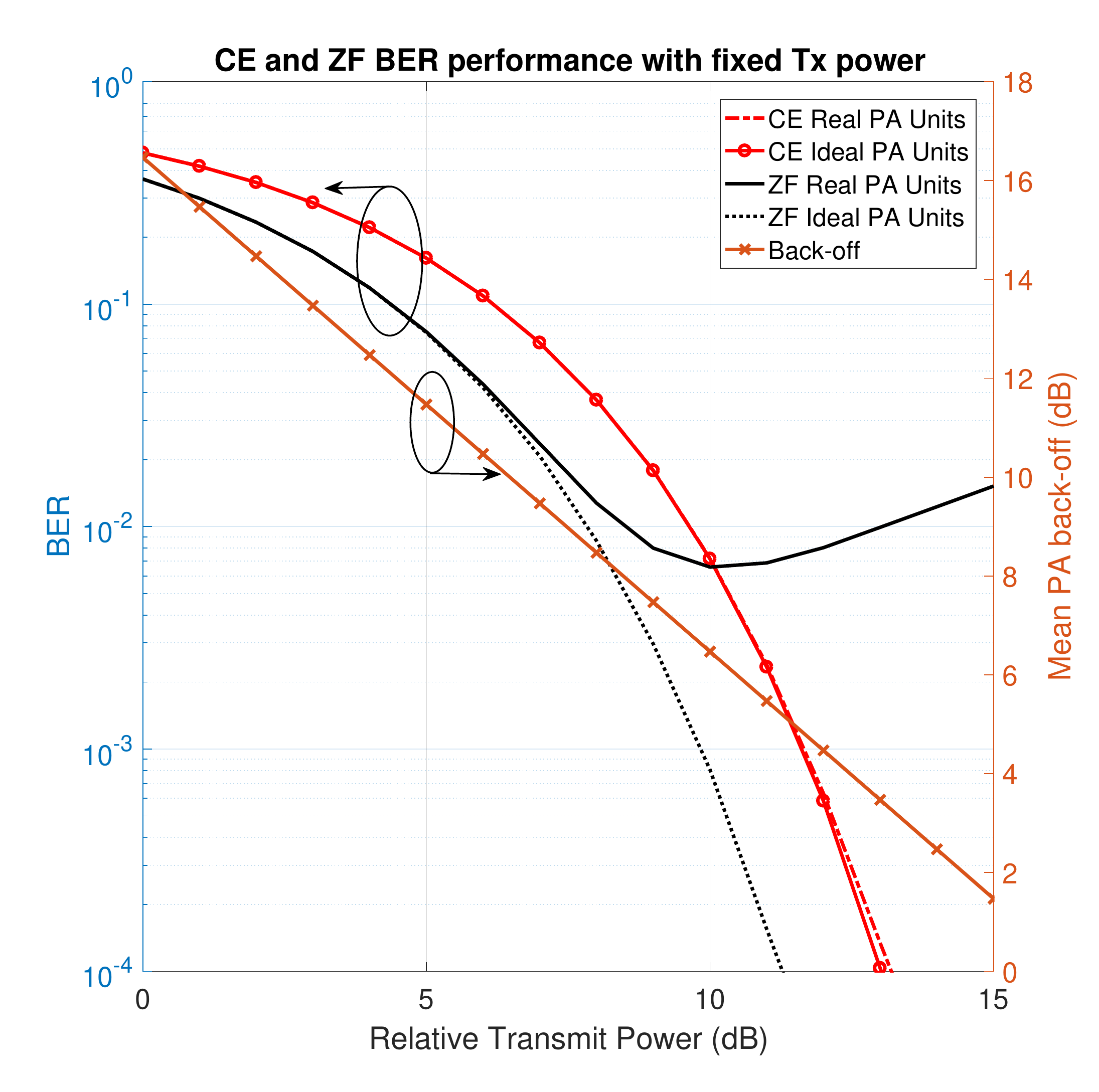}
	\caption{BERs of CE and ZF precoders with (i) \textit{Ideal PA Units} referring to a case with fully linear PAs and with (ii) \textit{Real PA Units} referring to the case with actual measured nonlinear PAs. Precoders are scaled such that both yield the same Tx power. Back-off ranging between 1.5 and 16.5 dB. Relative transmit power of 0 dB corresponds to receiver thermal noise SNR of -1 dB in case of CE precoder, and 0.9 in case of ZF precoder due to the larger beamforming gain.}
	\label{fig:si_beam_fig}
\end{figure} 
 The results are shown in Fig. \ref{fig:si_beam_fig} where it can be observed that for low transmit powers, ZF precoder outperforms CE by 1.9 dB due to the larger beamforming gain. However, when the effect of in-band distortion due to the nonlinear PAs start to become larger, the actual benefit of CE precoder becomes again evident, exhibiting a big gain at higher transmit powers, of around 5 dB, compared to the traditional ZF precoder.
\subsection{SINR Characteristics}
\begin{figure}
	\centering
	\includegraphics[width = 0.45\textwidth]{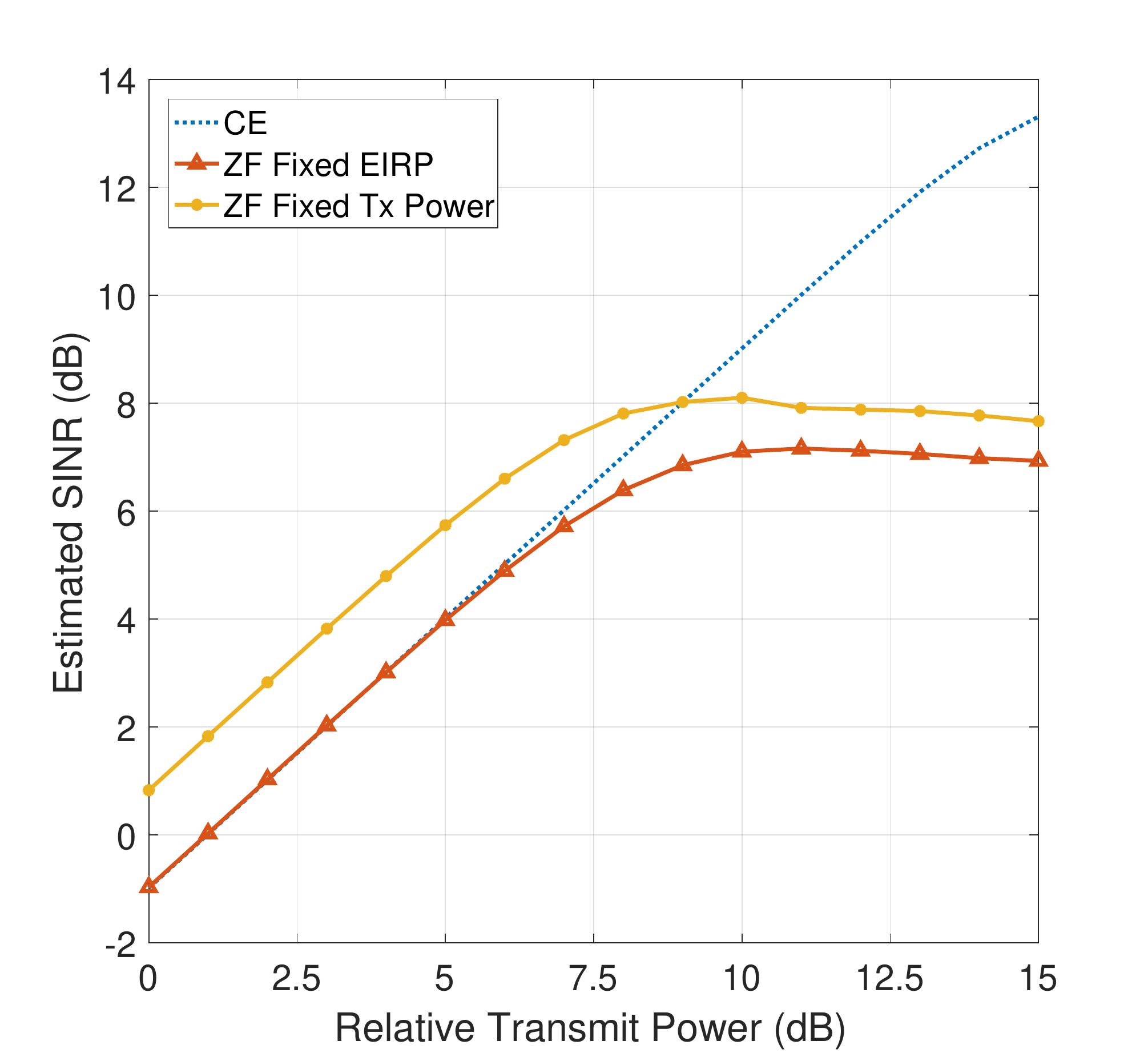}
	\caption{SINRs of CE and ZF precoders with (i) \textit{ZF Fixed Tx Power} referring to the case in which the transmit sum-power of ZF is the same as that of CE precoder, and with (ii) \textit{ZF Fixed EIRP} referring to the case in which ZF precoder has the same EIRP as that of CE precoder.}
	\label{fig:sinr_estimation}
\end{figure}

In Fig. \ref{fig:sinr_estimation}, we show how the relative transmit powers map into SINR at the receiver side. Since we assume perfect CSI knowledge, ZF precoder is capable of fully suppressing the MUI, while the CE precoder is designed and optimized to guarantee a minimum of 20 dB suppression (20 dB signal-to-MUI ratio). Such level of 20 dB MUI suppression can be safely assumed to be sufficient in most practical receiver scenarios, as the thermal noise SNR in cellular systems is commonly less than 20 dB and thus the MUI is below the thermal noise floor. 
From Fig. \ref{fig:sinr_estimation} we can conclude that the PAs exhibit very linear performance when fed by the CE precoded waveforms. The SINR increases linearly relative to the linear increase in the transmit power. It can also be seen that the transmit power maps into useful signal plus negligible in-band distortion, although for 15 dB of relative transmit power it exhibits a small reduction in the slope. On the other hand, in the case of ZF precoded signal, the performance is much worse. The SINR clearly exhibits a saturation behaviour due to the substantial in-band distortion suffered by the ZF precoded waveforms when passing through the PAs. We can also observe that the CE precoder performance is, at best, 5 dB above that of ZF, which is a considerable gain.

In order to obtain further insight of the obtained results, one can differentiate between two scenarios. First, a scenario where the TX power is relatively low, and thus, the resulting in-band distortion exhibited by the ZF precoder is not sufficiently large to allow CE precoder to outperform ZF precoder. Second, a high transmit power scenario in which CE precoder outperforms ZF precoder due to the increasing effect of the nonlinearities of the PAs. Such observation leads us to consider employing different precoders depending on the adopted transmit power in order to provide a better overall system performance. Since CE precoder provides somewhat lower beamforming gain, it is not adequate for transmit powers that do not allow to take advantage of the PAPR reduction. However for the second scenario, a CE precoder would allow to increase the performance significantly.

\section{Conclusion}
In this article, we studied constant envelope (CE) like precoding in multiuser large-array or massive MIMO systems. First, a computationally efficient optimization approach to obtain such CE precoded antenna samples was formulated. Then, the achievable multiuser radio link performance was addressed and analyzed under the effects of practical measurement-based nonlinear PA units in the transmitting array, using the CE precoder as well as the well-known ZF precoder for reference. The analysis and evaluations showed that despite providing around 2 dB lower beamforming gain than the ZF precoder, the PAPR reduction of the CE precoder is sufficiently large to allow it to outperform ZF in multiuser radio link performance at high transmit powers, i.e., when the PAs are pushed towards their saturating region. The actual SINR gain exhibited by CE precoder was shown to be up to 5-6 dB under realistic assumptions. Furthermore, the two highlighted transmit power scenarios, low and high, lead us to consider the adoption of a transmit power aware precoding approach, such that the CE precoder is deployed when the used transmit power allows to take advantage of the PAPR reduction, while ZF precoder can be then adopted at lower power levels. 

\section*{Acknowledgment}
This work was financially sponsored by the Academy of Finland (under projects 288670 and 301820) as well as by the Finnish Funding Agency for Innovation (Tekes), Nokia Bell Labs, Huawei Finland, Qualcomm, Pulse Finland and Sasken Finland under the project 5G TRX. The work was also supported by TUT President Graduate School, and by Tekes under the TAKE-5 project.


\begin{thebibliography}{99}

\bibitem{Green} M. Olsson, C. Cavdar, P. Frenger, S. Tombaz, D. Sabella and R. Jantti, "5GrEEn: Towards Green 5G mobile networks," \textit{2013 IEEE 9th International Conference on Wireless and Mobile Computing, Networking and Communications (WiMob)}, Lyon, 2013, pp. 212-216.
\bibitem{pa} V. Mancuso and S. Alouf, "Reducing costs and pollution in cellular networks," \textit{IEEE Communications Magazine}, vol. 49, no. 8, pp. 63-71, August 2011.
\bibitem{problemas_papr} Seung Hee Han and Jae Hong Lee, "An overview of peak-to-average power ratio reduction techniques for multicarrier transmission,"  \textit{IEEE Wireless Communications}, vol. 12, no. 2, pp. 56-65, April 2005.
\bibitem{oportunidades} F. Rusek et al., "Scaling Up MIMO: Opportunities and Challenges with Very Large Arrays," \textit{IEEE Signal Processing Magazine}, vol. 30, no. 1, pp. 40-60, Jan. 2013.
\bibitem{CE1} S. K. Mohammed and E. G. Larsson, "Per-Antenna Constant Envelope Precoding for Large Multi-User MIMO Systems,"  \textit{IEEE Transactions on Communications}, vol. 61, no. 3, pp. 1059-1071, March 2013.
\bibitem{CE2} S. K. Mohammed and E. G. Larsson, "Constant-Envelope Multi-User Precoding for Frequency-Selective Massive MIMO Systems,"  \textit{IEEE Wireless Communications Letters}, vol. 2, no. 5, pp. 547-550, October 2013.
\bibitem{CE3} J. Pan and W. K. Ma, "Constant Envelope Precoding for Single-User Large-Scale MISO Channels: Efficient Precoding and Optimal Designs,"  \textit{IEEE Journal of Selected Topics in Signal Processing}, vol. 8, no. 5, pp. 982-995, Oct. 2014.
\bibitem{CE4} A. Liu and V. K. N. Lau, "Two-Stage Constant-Envelope Precoding for Low-Cost Massive MIMO Systems,"  \textit{IEEE Transactions on Signal Processing}, vol. 64, no. 2, pp. 485-494, Jan.15, 2016.
\bibitem{CE5} J. C. Chen, C. K. Wen and K. K. Wong, "Improved Constant Envelope Multiuser Precoding for Massive MIMO Systems,"  \textit{IEEE Communications Letters}, vol. 18, no. 8, pp. 1311-1314, Aug. 2014.
\bibitem{CE6} S. Mukherjee and S. K. Mohammed, "Constant-Envelope Precoding With Time-Variation Constraint on the Transmitted Phase Angles,"\textit{ IEEE Wireless Communications Letters}, vol. 4, no. 2, pp. 221-224, April 2015.
\bibitem{ZF_c} C. B. Peel, B. M. Hochwald and A. L. Swindlehurst, "A vector-perturbation technique for near-capacity multiantenna multiuser communication-part I: channel inversion and regularization,"  \textit{IEEE Transactions on Communications}, vol. 53, no. 1, pp. 195-202, Jan. 2005.
\end{thebibliography}
\end{document}